\documentclass[twocolumn,showpacs,preprintnumbers,amsmath,amssymb]{revtex4}

\usepackage{graphicx}
\usepackage{dcolumn}
\usepackage{bm}

\newcommand{\be}{\begin{equation}}
\newcommand{\ee}{\end{equation}}
\newcommand{\eq}[1]{(\ref{#1})}

\def\nn{\nonumber}
\def\bea{\begin{eqnarray}}
\def\eea{\end{eqnarray}}

\def\bk{{\bf k}}

\def\bx{{\bf x}}

  \def\cX{{\cal X}}

\def\Tr{{\rm Tr}}
\def\one{\mbox{1 \kern-.59em {\rm l}}}

\begin{document}

\preprint{hep-th/0602148}
\preprint{RIKEN-TH-65}

\title{Non-local Matching Condition and\\
Scale-invariant Spectrum in Bouncing Cosmology}

\author{Chong-Sun Chu}
\affiliation{Centre for Particle Theory 
and Department of Mathematics,
University of Durham, Durham, DH1 3LE, United Kingdom}
\email{chong-sun.chu@durham.ac.uk}


\author{Ko Furuta}%
\affiliation{Theoretical Physics Laboratory,
The Institute of Physical and Chemical Research (RIKEN),
2-1 Hirosawa, Wako, Saitama 351-0198, Japan}
\email{furuta@riken.jp}

\author{Feng-Li Lin}
\affiliation{Department of Physics,
 National Taiwan Normal University,
 Taipei, Taiwan 116}
\email{linfengli@phy.ntnu.edu.tw}
\date{\today}

\begin{abstract}
In cosmological scenarios such as the pre-big bang
scenario or the ekpyrotic scenario, a matching condition between the
metric
perturbations in the pre-big bang phase and those in the post big-bang
phase is often assumed. Various matching conditions have been
considered in the literature. Nevertheless obtaining 
a scale invariant CMB spectrum via a concrete mechanism
remains impossible.
In this paper, we examine this problem from the point of view of local
causality. We  
begin with introducing the notion of local causality and explain
how it constrains the form of the matching condition. We then prove a
no-go theorem: independent of the  details of the matching
condition, a scale invariant spectrum is impossible as long as the local
causality condition is satisfied. 
In our framework, it is easy to show that a violation  of local
causality around the bounce is needed in order to give
a scale invariant spectrum. We study a specific scenario of this
possibility by considering a nonlocal effective theory inspired by
noncommutative geometry around the bounce and show that a scale
invariant spectrum is possible. 
Moreover we demonstrate that
the magnitude of the spectrum is compatible with
observations if the bounce is assumed to occur  at an energy scale
which is a few orders of magnitude below the Planckian energy scale.
\end{abstract}

\pacs{Valid PACS appear here}
\maketitle

\section{\label{sec:level1}Introduction
}

The studies of big bang singularity is an important arena where string
theory and cosmology meet. Recently based on stringy dualities or extra
dimensions arguments, attempts to resolve 
the big bang
singularity such as the pre-big-bang \cite{pbb} or the ekpyrotic/cyclic
cosmology \cite{ekpyrotic},\cite{cyclic} has been put forwarded
(see for example \cite{transp} for 
review on string cosmology).
To be an ambitious model of string cosmology,
one hopes to reproduce the scale-invariant spectrum of
density perturbations without invoking inflation. In both
of these scenarios, 
one can expect the  physics around the bounce (say around 
$\eta_-<\eta <\eta_+$) to be nonperturbative and  
highly nontrivial.
Without committing oneself to any specific form of the
dynamics involved, a useful approach to this problem is to replace the
dynamical evolution around the bounce by a nontrivial phase
transition \cite{DerMuk,HwVish}.
In this approach, one evolves the Einstein equation far from the
bounce (for time $\eta <\eta_-$ and $\eta >\eta_+$) where the classical
equation can be applied, and then try to connect the physics at $\eta_-,
\eta_+$ using an appropriate matching condition. 
We call these matching
conditions to distinguish from the usual junction conditions
\footnote{
In general relativity, junction conditions are used to match 
solutions to the
equation of motion across a surface of discontinuity, for example, one
caused by a sharp localization of matters. Some well-known junction
conditions are the Lichinerowicz junction condition, the
Isarel junction condition and the O'Brien-Synge junction
condition \cite{jun}. 
However in studies such as the bounce cosmologies
where general relativity may break down around the bounce, 
junction condition may prove limited and 
generalization such as the matching condition 
are needed. }. 

The form of the matching conditions 
has important consequences on the form of the cosmic microwave
background (CMB) spectrum. 
In the framework of known physics, it is found 
not to be the case both for
the pre-big bang \cite{Brustein:1994kn,DurrVer} and 
the ekpyrotic scenarios
\cite{DurrVer,Lyth:2001pf,Brand,Cremin}.
%
%
It is natural to ask to what extent this conclusion depends on the details
of the matching conditions. 
In \cite{Cremin} it was argued that the
predictions of density perturbations of the bouncing cosmology are
independent of the details of the matching condition near the bounce,
with the
reason that there exists a dynamical attractor such that every
observer will follow the same history and result in the continuity of
the curvature perturbation.
Studies on models with regular bounce has been
carried out \cite{Allen:2004vz} and gave negative result
on rather general assumptions \cite{Bozza:2005qg}.

The main goal of this paper is to understand to what extent the known
matching conditions may get modified due to the high energy corrections
to Einstein gravity and to examine whether or not there is any special
class of matching conditions that could lead to a scale invariant
spectrum. We will show that if the matching condition respects a local
causality condition near the bounce, there is indeed no way to generate
a scale invariant power spectrum, agreeing with the general result of
\cite{Cremin}. By local causality, we mean a local event is not allowed
to affect infinitely separated points through the bounce. 
More specifically, using our local matching condition, 
we find that a new mixing
term between the subdominant mode in the pre-bounce era and the
subdominant mode in the late time universe is allowed. This could be
attributed to the anisotropic stress during the bounce. This new mixing
term, however, cannot help to produce a scale-invariant mode that is
dominant in the late time universe. To do this job, a mixing between the
dominant mode in the pre-bounce era and the dominant mode in the late
time is needed. However, this mixing is absent in general if the
matching condition respects the local causality condition. 

The important conclusion of our general treatment of the matching
condition is that there has to be new nonlocal effects beyond
general relativity  in order  to obtain 
a scale invariant power spectrum 
for the CMB fluctuations.
In string theory,
nonlocal effects such as noncommutative geometry has been widely
considered. Motivated by this, we consider a modified equation of motion
for the cosmological fluctuations inspired by the noncommutative field
theory. We find that if the degree of nonlocality is not too strong,
a scale invariant spectrum may appear.
Moreover the magnitude of the density spectrum is compatible with
observation if the bounce occurred at an energy scale that is 
a few orders of magnitude below the Planck scale.

The paper is organized as follows. In the next section, we discuss the
notion of local causality and explain how it leads to the condition that
there exists no negative powers of derivative in the matching condition.
In section \ref{sec:level3}, 
we begin with a brief review of the linear cosmological
perturbation theory, applied to the pre-big bang and ekpyrotic
scenarios. Next we impose the condition of local causality and express
our local matching condition in terms of the gauge invariant variables.
We then demonstrate 
a no-go theorem: a scale invariant spectrum is generally 
impossible if a local matching condition is assumed. 
In section \ref{sec:level4}, we go beyond the condition of local causality 
and consider a toy model inspired by the noncommutative field theory
of the ekyprotic scenario. We find in this
model that it is possible to obtain a scale-invariant spectrum.
Section \ref{sec:level5} is devoted to conclusions and discussions.
The paper is ended with a couple of appendices which address some of the
more technical issues.

Here is our notation: $\mu, \nu,\cdots$ run from $0$ to $3$, $i,j,\cdots$ 
run from $1$ to $3$, ${\bf x}$ represents the spatial 
part of the coordinate $x^{\mu}$.


\section{\label{sec:level2}
Local matching condition in bouncing cosmologies}

As we explained in the introduction, the situation is more complicated
in the studies of bouncing cosmologies. Here one cannot evolve the
classical general relativity in the region around the bounce since it is
supposed to break down there. Since this region is supported generally
on a nonzero measure set, one cannot apply the usual idea of junction condition 
which is imposed at a single hypersurface. One need a more
general matching condition. 
 
One approach which has been widely adopted in the literature is to
assume the occurrence of a nontrivial phase transition around the bounce.
Let ${\cal O}(t, \bx)$ 
be the order parameter associated with
the phase transition. Then the matching is carried out on the two
hypersurfaces  ($\pm$ corresponds to initial and final time of the
phase transition)
\be
{\cal O}(t_\pm, \bx) ={\cal O}^0_\pm ={\rm constant},
\ee
by matching up the values of certain quantity which one may argue to be
conserved during the phase transition. However there is weakness in
this approach since the conservation law depends in some details of the
dynamics during the phase transition. 

In the following we will take a different approach which allow us to
discuss  all possible matching conditions in general. We will
characterize a matching condition according to whether it respects 
a local causality condition of the physics involved.  
Our approach has the advantage that it is more robust 
without assuming the underlying dynamics.  
Also as we will see, it gives
clear indication 
how new physics should appear in order to
give a scale invariant density spectrum.

\subsection{\label{sec:level2_1}
Requirement of local causality and local matching condition}

Our guiding principle in constructing the matching condition is the
condition of local causality. The statement is that no local event is
allowed to affect infinitely separated points through the phase
transition. 
Or equivalently the past causal region of any space-time point is 
of finite extent for a finite time duration.
Although this
assumption is intuitive and natural, a couple of remarks about its applicability
is in order.
\\
1. We note that if the spatial section of the universe during the bounce is
compact and of the size comparable to the time scale of the 
bounce, local causality is not expected to hold. 
This would be the case
if the size of the universe is the self dual radius of the string theory
during the bounce. 
Several models with closed spatial section  
has been studied \cite{Peter:2004um}.
On the other hand, if the spatial section of the
universe during the bounce remained 
of the size much larger than the time
scale of the bounce, it is reasonable to 
assume the validity of local causality.
This requires the flattness problem to be solved by some mechanism prior
to the bounce \cite{Lindeetal}.
We do not address the flattness problem in this paper and 
concentrate on the problem of the spectrum of the density fluctuation.
\\
2. Due to the nonlocal nature of string, 
modification of the local causality
condition occurs, typically at the interacting string level \cite{causal}. One
can expect the violation of the local causality condition also occurs in
some stringy cosmologies. Another well known example of nonlocal physics
is noncommutative geometry. It is possible that quantum gravity may lead
to a description of noncommutative geometry during the bounce and 
a violation of the local causality condition. We will say more about this
later.

Next  we want to
apply the requirement of the local causality 
to obtain our local matching condition. Let us begin with introducing 
two space-like hypersurfaces $S_-$ and $S_+$ separated by the bounce,
where the former is 
before the bounce and the latter is after it.
Before and after $S_\pm$, 
we assume that general relativity is valid
and that linear perturbation provides a good approximation for the
evolution of the fluctuation. 
The surface of the matching is specified by equations
\begin{equation}
{\cal O}_{\pm}(t,\bx)={\cal O}^0_{\pm} \;\mbox{ on } S_{\pm}
\label{surf_pm}
\end{equation}
respectively.
Here ${\cal O}_{\pm}(t,\bx)$ 
are scalar quantities
constructed out of metric and matter fields and ${\cal O}^0_{\pm}$
are constants.
Let us choose the synchronous gauge
\begin{equation}
g_{00}=-1,\quad g_{i0}=g_{0i}=0,
\label{sync}
\end{equation}
We also choose the time coordinate to satisfy 
\begin{equation}
t=t_{\pm}\; \mbox{ on }S_{\pm}
\label{t_pm}
\end{equation}
respectively. Note that the condition (\ref{sync})
and (\ref{t_pm}) are 
possible for any surface of the matching
(\ref{surf_pm}). This is due to the residual gauge
degrees of freedom in the syncronous gauge.

We will be interested in the standard type of 
cosmological backgrounds which are homogeneous and isotropic, 
\begin{equation}\label{bmetric}
ds^2=-dt^2+a^2d\bx^2
=a^2(-d\eta^2+d\bx^2).
\end{equation}
Here $a= a(\eta)$ and $\eta$ is the conformal time.
Consider metric and scalar field perturbation of the form
\begin{equation} \label{pert}
\delta g_{\mu\nu}
= \delta\tilde{g}_{\mu\nu}(k,\eta) e^{i\bk\bx} 
\;\mbox{ and  }\;
\delta \varphi= \delta\tilde{\varphi}(k,\eta) e^{i\bk\bx},
\end{equation}
where $k$ is the comoving wave number. 
We will be interested in 
perturbations with  wavelength longer than
the horizon size 
since these are the fluctuations 
that are relevant to the CMB observational data. 
The long wavelength limit means that the physical wavelength $a/k$ 
is much larger than the Hubble scale, i.e.
\begin{equation}
\frac{k}{aH}\bigg|_{\eta=\eta_{\pm} }
=\frac{k}{\mathcal{H}}\bigg|_{\eta=\eta_{\pm} }
\ll 1.
\label{longwavecond}
\end{equation}
Here $H={\dot{a}}/{a}$ is the Hubble constant and
$\mathcal{H}=a'/a$; dot represents the derivative with 
respect to cosmological time $t$ and prime
represents the derivative with respect the conformal
time $\eta$.
One can convince oneself that this bound for long wavelength limit is
very easy to be satisfied by the currently observed  CMB data.

\begin{figure}[htbp]
\begin{center}
\includegraphics[scale=.7]{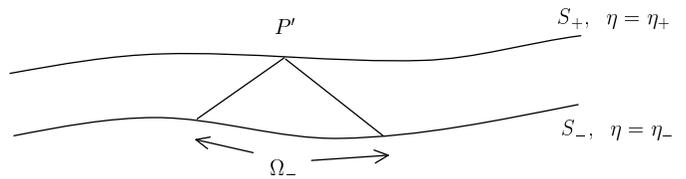}
\caption{\label{fig:1} Local cauality}
\end{center}
\end{figure}

Now we impose the condition of local causality.
By local causality of the matching condition we means that there exists
some finite region $\Omega_-$ on $S_-$ such that all the data
$\{ g_{ij}, g'_{ij}, \varphi, \varphi'\}$ at $P'$ on $S_+$
is determined only by data $\{g_{ij}, g'_{ij}, \varphi,
\varphi'\}$ on $\Omega_-$, 
modulo local spatial coordinate
transformation (see FIG. \ref{fig:1}).
A subtle point in the problem of the matching condition 
is that it is not guaranteed that 
the Cauchy problem is solvable throughout the bounce.
In this paper, we assume the solvability for long 
wavelength modes.
In the long wavelength limit, 
the scalar field and metric on 
$\Omega_-$ 
can be expanded 
in terms of $\varphi$, $g_{ij}$
and derivatives of them at a point $P$ on $\Omega_-$.
Given a
one to one map $P\to P'$ from $S_-$ to $S_+$,
the metric and scalar field at $P'$ is given by a function of 
the scalar field, metric and derivatives of them at $P$.
For example, once the coordinate system on $S_+$ is given,
the matching condition for the metric formally 
takes the form
\begin{equation} 
g_{ij}\big|_{\eta= \eta_+,P'}=H_{ij}(g_{ij},g_{ij}',\varphi,
\varphi', \nabla_k
)\big|_{\eta= \eta_-,P} \;\; , \label{matchingex}
\end{equation}
The function $H_{ij}$ is a rank 2 symmetric tensor with respect to 
the spatial coordinate transformation.

Because of the local causality, no negative power of
$\nabla_k$ can appear. 
Due to the tensor structure, it is clear that
only even powers of $\nabla_k$ appear.
The general form of the metric on $S_+$ is
given by possible covariant combinations of metric and scalar fields
with spatial derivatives on $S_-$.
Choosing a map $P\to P'$ properly, no
$\bx$-dependent quantity
enters into the matching condition.
Expanded in powers of $\nabla_k$, 
the matching condition \eq{matchingex}
takes the form
\be\label{matchingho}
 g_{ij}\big|_{\eta=\eta_+,P'}
=
H^{(0)}_{ij}(g_{ij}, g'_{ij}, \varphi, \varphi') 
\big|_{\eta=\eta_-,P} +  \cdots , 
\ee
where $\cdots$ denotes terms of higher order in $\nabla_k$. 
Now note that the 
only nontrival tensors available for constructing the tensor 
$H^{(0)}_{ij}$ are $g_{ij},g_{ij}'$ and 
$g^{ij}, g'{}^{ij}$. 
Employing a matrix notation, 
it is easy to see that the most general form of  $H^{(0)}$ is:
\be \label{H0}
H^{(0)} =  \sum_\cX c_\cX(\varphi,\varphi') P_\cX(g, g', g^{-1},
g'{}^{-1}),
\ee
where for $n\geq 0$ integer,
\begin{align} \label{Pchi}
&P_\cX := f(AB) A (BA)^n  \nonumber\\
 &\mbox{with $A=g$ or $g'$,\;
$B=g^{-1}$ or $g'{}^{-1}$}.
\end{align} 
$f(AB)$ represents all possible functions constructed out of
$\Tr(AB)^l$'s.
The sum
$\cX$ in \eq{H0} is over all possible inequivalent forms
of such factors. 
Similarly, we can generalize the above discussion to the matching condition of 
$g'_{ij}$ and for other generic tensors. 

We remark that one may try to generalize the above to the
gauge invariant quantities, but the requirement of local causality is
less clear and less easy to formulate since spatial integrals of the
metric fluctuaions
is included in the definition of the gauge invariant quantities.
Our local matching condition is in the same spirit 
as the method of spatial gradient 
expansion (see, e.g., \cite{Parry:1993mw}).
However, validity of a four dimensional effective theory
during the bounce is not assumed in our approach.
In the next section we will apply our local matching condition 
\eq{matchingho} 
to the problem of bouncing cosmologies.


\section{\label{sec:level3}
Density perturbation in  bouncing cosmologies}

In this section, we will first briefly review the theory of cosmological
perturbation \cite{MFB}. We then apply the 
local  matching condition to the bouncing cosmologies.
Before the bounce, the model for the pre-big bang or 
ekpyrotic scenario is given by the four-dimensional effective theory
with a scalar field coupled to gravity, the action is
\begin{equation}
S=\int d^4 x\sqrt{-g} \left( \frac{1}{2\kappa^2}R+\frac{1}{2}
\left(\partial \varphi\right)^2-V(\varphi)\right).\label{scalarmodel}
\end{equation}
The scalar field $\varphi$ is related to dilaton 
or the size of the extra-dimension.
The evolution of fluctuation depends on the nontrivial 
potential $V(\varphi)$ for the scalar field. 
The universe starts from a Minkowski spacetime
with $V\approx 0$. 
In this paper we will consider the case of having a single scalar 
$\varphi$ only. The generalization to many scalars is straightforward.  

\noindent We remark that:
1. In general there could be other fields in the model. 
One can divide the scalar fields
into the  background one (adiabatic field) and the others (entropy fields)
\cite{Gordon:2000hv}. 
The scalar field in (\ref{scalarmodel})
is the adiabatic field and entropy fields are zero
by definition. The effect coming from fluctuation in 
the other scalar fields contribute to the 
entropy perturbation. We will discuss the effect of
the entropy perturbation on the matching condition in 
the appendix \ref{sec:levelA_C}. 
\\
2. In addition to the local causality condition, we will also assume the 
perfect fluid condition before and after the bounce.
Perfect fluid condition requires the vanishing
of the anisotropic stress. This is a reasonable assumption
since in the single scalar field model
(\ref{scalarmodel}), there is no anisotropic  
stress in the linear order
of perturbation theory.
In the radiation dominated era after the bounce, 
there is also no anisotropic stress if the mean free path is small, as
is often assumed. We will discuss the effect of the anisotropic stress
in the appendix \ref{sec:levelA_D}. 

In section \ref{sec:level3_3}, we will show that one cannot obtain a scale invariant
spectrum if the local matching condition is employed. This conclusion is
unaffected by the effects discussed in appendix \ref{sec:levelA_C} and 
\ref{sec:levelA_D}.

\subsection{\label{sec:level3_1}
Review of cosmological perturbation  theory}

Metric fluctuations are classified by their properties
under the spatial rotations into scalar, vector 
and tensor ones. In the linear perturbation theory, time evolution of
these perturbations are decoupled with each other. In this paper, 
we will  focus on the scalar ones (the consideration of the vector and
tensor modes can be proceeded similarly) which are defined by 
\begin{equation}\label{def-scalar-fluct}
\delta g_{\mu\nu}=a^2 
\begin{pmatrix} 
2\phi & -B_{|i} \\ 
-B_{|i} & 2(\psi \delta_{ij} -E_{|ij})
\end{pmatrix},
\end{equation}
where $\phi$, $B$, $\psi$ and $E$ are scalar functions on the constant-time 
hypersurface. 
Indices $|i$ means taking three dimensional covariant derivative 
in spatial direction. 
In our case it is the ordinary derivative.
Indices $i,j\cdots$ are raised and lowered by the Kronecker delta.

There is a gauge symmetry for the fluctuations 
of metric due to the residual diffeomorphism on the background metric,
and the gauge transformations affecting scalar fluctuations are 
of the form.
\begin{equation}
\eta \to \tilde{\eta}=\eta + \xi^0 (\eta,\bx)
\hspace{3mm}\mbox{ and }
x^i \to \tilde{x}^i=x^i+\delta^{ij}
\partial_j \xi(\eta,\mbox{\boldmath{x}}).
\label{gaugetr}
\end{equation}
In the linear order, 
functions $\phi$, $B$, $\psi$ and $E$ in the metric fluctuation
are changed to
$\tilde{\phi}$, $\tilde{B}$, $\tilde{\psi}$ and $\tilde{E}$
as follows.
\begin{equation}
\begin{split}
\tilde{\phi}=\phi-(a'/a)\xi^0 -{\xi^0}',& \hspace{6mm}
\tilde{\psi}=\psi+(a'/a)\xi^0, \\
\tilde{B}=B+\xi^0-\xi' ,& \hspace{6mm} \tilde{E}=E-\xi.
\end{split}
\label{flucttransf}
\end{equation} 
Gauge invariant quantities are 
\begin{equation}
\Phi=\phi+(1/a)[(B-E')a]', \quad
\Psi=\psi-(a'/a)(B-E').
\label{defPhi}
\end{equation}
The energy momentum tensor of the action (\ref{scalarmodel}) describes
a perfect fluid.
It follows that $\Phi=\Psi$. 
We will assume that the perfect fluid condition is satisfied for 
$\eta<\eta_-$ and $\eta>\eta_+$.

For the fluctuation of the scalar field we have the constraint 
equation coming from the equation of motion,
\begin{equation}
\delta\varphi^{(gi)}=\left(\frac{3}{2}l^2\varphi'\right)^{-1}
(\Psi'+\mathcal{H}\Phi) ,
\label{phivarphi}
\end{equation}
where 
\begin{equation}
\delta\varphi^{(gi)}:=\delta\varphi+\varphi'(B-E')
\end{equation}
is the gauge invariant fluctuation of the scalar field
and $l$ is the Planck length.
Thus the scalar fluctuation $\Phi$ is the only 
independent quantity to consider.
By expanding the Einstein equation around the  
background metric, one gets the time-dependent equation for $\Phi$, 
which can be put into
a compact form in terms of the new variable
$u := a \Phi /\varphi'$:
\begin{equation}
{u}''(k,\eta)+\left(k^2-\frac{(z^{-1})''}{z^{-1}}\right)u(k,\eta)
=0,
\label{schroedinger}
\end{equation}
where $z={a\varphi'}/{\mathcal{H}}$.
This is a Schr\"{o}dinger type equation and can be 
solved once the initial condition
is specified. The initial condition is given by assuming that the
universe starts from the Minkowski space where  
$k^2 \gg{(z^{-1})''}/{z^{-1}}$. 
In this limit, we get a plane wave solution
with the normalization fixed by the canonical quantization, the result
is 
\begin{equation}
u(k,\eta)=-\frac{3}{2}l^2 k^{-3/2}e^{-ik\eta}.
\label{initial}
\end{equation}
Note that only the positive energy mode ($e^{-ik\eta}$) is chosen as
the initial condition.
 
On the other hand, we are interested in the long wavelength limit,
$k^2\ll (z^{-1})''/z^{-1}$ near the bounce,  therefore
\be
{u}''(k,\eta) -\frac{(z^{-1})''}{z^{-1}} u(k,\eta)
=0.
\label{u-eq}
\ee
The solution at the
leading order of $k$-expansion is  
\begin{equation} 
l^{-2} u=S(k) 
\big(\frac{1}{z}+O(k^2)\big)
+\frac{3}{2} D(k) \big( \frac{1}{z}\int d\eta z^2
+O(k^2)\big)\label{u}.
\end{equation}
Note that the correction starts from
$k^2$ order. From (\ref{u}) we have
\begin{align} \label{Phiin k}
\Phi=&S(k) \left(l^2\frac{\mathcal{H}}{a^2}+O(k^2)\right)
\nonumber\\
&+D(k)\left(\frac{1}{a}\left(
\frac{1}{a}
\int d\eta a^2\right)'
+O(k^2)\right).
\end{align}
The coefficients $S(k)$ and $D(k)$ are time independent. 
Their form are fixed by extrapolating (\ref{u}) to the short
wavelength limit and compare with (\ref{initial}) at the regime
$k^2\sim (z^{-1})''/z^{-1}$. 
The details depend on the time evolution of  
$\Phi$ at this intermediate regime and thus depends on the potential 
$V(\varphi)$.
A scale invariant spectrum 
\begin{equation}
P_\Phi (k)=\frac{k^3}{2\pi^2}\left|\Phi(k)\right|^2 
\end{equation}
is obtained if $\Phi(k) \propto k^{-3/2}$.

In bouncing cosmologies, the term proportional to $S$ grows before the bounce
whereas the term proportional to $D$ is constant with respect to time. 
After the bounce, assuming the universe is filled with  
radiation  and  that there is
no entropy perturbation, then the evolution of fluctuation is 
governed 
by the same equation \eq{schroedinger} as before. 
In the long wavelength limit, the solution for the fluctuation
is in the same form as (\ref{Phiin k}).
We note that the $S$-mode is now decaying
in time and thus the $D$-mode becomes dominant, particularly
at the time of decoupling. The important task is to determine the
$k$-dependence of $S^{(+)}(k)$ and $D^{(+)}(k)$ 
after the bounce. Using the matching
condition, one can relate them to the $S^{(-)}(k)$ and $D^{(-)}(k)$ 
before the bounce.

In general a mixing between the modes $S$ and $D$ may occur,
\begin{equation} \label{DDS-mix}
\left(
\begin{array}{c}
S^{(+)}\\
D^{(+)}
\end{array}
\right)
= M \left(
\begin{array}{c}
S^{(-)}\\
D^{(-)}
\end{array}
\right).
\end{equation}
Here, matching matrix $(M)_{ij}$  
are functions of $k$.  If the mixing is right,
the desired form of $D^{(+)} \sim k^{-3/2}$ may be generated and hence
result in a scale invariant spectrum in CMB. 

Below we derive the form
of the matching between $(S^{(-)}, D^{(-)})$ and $(S^{(+)}, D^{(+)})$.
In particular we will find that local causality implies that
\be
M_{21} =0, \quad M_{22} =1, \quad \mbox{and thus}\quad D^{(+)} = D^{(-)}.
\ee

\subsection{\label{sec:level3_2}
Consequences of the local matching condition: mixing matrix}

Our discussion so far is general and  
we can consider the matching condition on any surface.  
In the analysis below, we will take the matching surface $S_-$ to be
the one given  by $\varphi=\mbox{constant}$. 
We can choose a different matching surface
with ${\cal O}_-(\eta,x)=\mbox{constant}$ for a different quantity
${\cal O}_-$.
However as explained in the appendix \ref{sec:levelA_B}, 
the metric and scalar field fluctuations
turns out to be  the same as 
(\ref{gconstphi}) and (\ref{phiconstphi}) up to the
order we consider and leads to the same matching 
condition (\ref{general}) below . 

To justify the linear matching condition, one needs to examine
the magnitude of the fluctuation of physical quantities on 
the surface of the matching. In \cite{Cremin},
validity of the linear perturbation theory is demonstrated
using the attractor property of the background solution.
In particular it has been shown that in 
the syncronous gauge  $\delta g_{00}=\delta g_{0i}=0$, 
linear perturbation theory
remains valid near the bounce. Another advantage of using 
the syncronous gauge is that we can choose the matching surface such
that the scalar field has no fluctuation (up to the order of $k$ we consider)
over it. 
In fact as we show in the appendix \ref{sec:levelA_A}, 
we can exploit the residual gauge symmetry of this 
gauge and choose the  hypersurfaces $S_-$ 
such that $\varphi$ is constant over $S_-$, i.e. $\delta \varphi =0$.
Therefore we will use the syncronous gauge.
The form of
the metric and scalar field fluctutaion in this gauge is 
also computed in
the appendix \ref{sec:levelA_A}
with the result that, 
around $\eta =\eta_-$,
\begin{widetext}
\be 
\delta g_{\mu\nu}^{(-)}(\eta,\bk)
=2a^2\left(
\begin{array}{cc} 
0 \;\;\;& 0 \\ 
0 \;\;\;& 
\begin{array}{c}\Big( D^{(-)}+O(k^{n_\Phi+2}))\Big) \delta_{ij}\\
+ k_i k_j \Big( S^{(-)}l^2\int d\eta a^{-2} 
- D^{(-)} \int d\eta(a^{-2} \int d\eta a^2)+O(k^{n_\Phi+2})
\Big)
\end{array}
\end{array}
\right).
\label{gconstphi}
\ee
\end{widetext}
and 
\be
\delta\varphi=O(k^{n_\Phi+2}).
\label{phiconstphi}
\ee
And such that $\delta \varphi =0$ at $\eta=\eta_-$.
Here $n_\Phi$ is a constant specifying the order in the wave number $k$
of the gauge invariant
observable $\Phi$, i.e. $\Phi=O(k^{n_\Phi})$.
In our following analysis, we will be interested in the leading order
terms only, and so any terms with $O(k^{n_\Phi+2})$ are
effectively zero in the long wavelength limit.
%
Note also that 
\begin{equation}
c_\cX=c_\cX^{(B)}+O(k^{n_\Phi+2}) \sim c_\cX^{(B)} 
\end{equation}
for the coefficient $c_\cX$' appearing in the general 
matching condition \eq{matchingho}.
Here the superscript  $B$ denotes the background.

As a first application of our local matching condition, 
we will  show that the covariance of the matching condition 
\eq{matchingho} proves the continuity through the bounce 
of the term proportional to $D^{(-)}\delta_{ij}$ 
in the metric fluctuation \eq{gconstphi}.
To see this, we substitute \eq{gconstphi} into the general 
matching condition
\eq{matchingho}
and, for the moment, assume that 
the term $k_i k_j S^{(-)} \cdots$  is higher order
in $k$ compared to the $D^{(-)}$ term.
Due to the structure of $P_\cX$, we get generally
\be
P_\cX(g, g', g^{-1},g'{}^{-1}) = d_\cX(a,a') \times (1+2D^{(-)}) +
\cdots
\ee
to leading order in $k$. Here $d_\cX(a,a')$ is a homogeneous monomial of
$a,a',a^{-1}, a'{}^{-1}$ of degree 1. 
Therefore we obtain 
\be
g_{ij}^{(+)}\big|_{\eta=\eta_+}=a_+^2(1+2D^{(-)})\delta_{ij}+ \cdots
\ee
where 
\be 
a_+= a_+(a(\eta),a'(\eta), \cdots)\big|_{\eta=\eta_-}.
\ee
Here
we have used the fact that $c_\cX^{(B)}$ depends only on $a$ and $a'$ and
we have denoted 
\be \label{aplus}
a_+^2 := \sum_\cX c_\cX^{(B)} d_\cX.
\ee 
As a result, we obtain
\begin{equation}
\delta g_{ij}^{(+)}\big|_{\eta=\eta_+}=2a_+^2D^{(-)}\delta_{ij}
+O(k^{n_\Phi+2}).
\end{equation}
Appling the same matching argument to the time derivative of the metric,  
we get
\begin{equation}
\delta g_{ij}^{(+)'}\big|_{\eta=\eta_+}=2(a_+^2)'
D^{(-)}\delta_{ij}+O(k^{n_\Phi+2})
\end{equation}
where the constant $(a_+^2)'$ is defined in a similar fashion as \eq{aplus}.
Having $\delta g_{ij}^{(+)}$ and its first time derivative  
on $\eta=\eta_+$ surface, then $\delta g_{ij}^{(+)}$
can be determined for $\eta>\eta_+$. In the synchronous gauge,
 we have 
\begin{equation}
\delta g_{\mu\nu}^{(+)}(\eta,\bk)
=2a_+^2\left(
\begin{array}{lc} 
0 & 0 \\ 
0 & 
D^{(-)}\delta_{ij}+O(k^{n_\Phi+2})
\end{array}
\right).
\label{metricafter}
\end{equation}
This concludes the continuity of the term proportional to
$D^{(-)}\delta_{ij}$ through the bounce.
This result is consistent with the argument given in \cite{Cremin}.

Next we include also the  $k_i k_j$ terms in the local matching 
condition \eq{matchingho}. The most general form
of $\delta g_{\mu\nu}^{(+)}$ after the bounce is
\begin{align}
&\delta g_{\mu\nu}^{(+)}(\eta,\bk)
=2a_+^2\nonumber\\
&\times\left(
\begin{array}{lc} 
0 \;\;\;& 0 \\ 
0 \;\;\;& 
\begin{array}{c}\big(D^{(-)}+O(k^{n_\Phi+2})\big)\delta_{ij}\\
+k_i k_j \Big(S^{(-)}M_S+
D^{(-)}M_D+O(k^{n_\Phi+2})\Big)
\end{array}
\end{array}
\right).
\label{metricafter1}
\end{align}
Here $M_S$ and
$M_D$ are some
functions of $\eta$ whose  form  can be fixed by requiring the
perfect fluid condition. Indeed from the definitions of the gauge invariant
quantities (\ref{defPhi}), 
one has
\begin{align}
\Phi=&(1/a)
\left[
\big(S^{(-)}M_S+
D^{(-)}M_D\big)'a\right]'
+O(k^{n_\Phi+2}),\label{Phi(ii)}\\
\Psi=&D^{(-)}-(a'/a)\big( S^{(-)}M_S+
D^{(-)}M_D\big)'
+O(k^{n_\Phi+2}).
\end{align}
Perfect fluid condition $\Phi=\Psi$ requires 
\begin{equation}
(1/a)[M_S'a]'=-(a'/a)M_S',
\label{S1}
\end{equation}
and
\begin{equation}
(1/a)[M_D'a]'=1
-(a'/a)M_D'.
\label{D1}
\end{equation}
Solving (\ref{S1}) gives
\begin{equation}
M_S'=C_S l^2/a^2,
\end{equation}
and (\ref{D1}) gives
\begin{equation}
M_D'=a^{-2}\left(\int d\eta a^2
+C_{D} l^2\right),
\end{equation}
where $C_S$ and $C_D$ are arbitrary constants.
Putting this into (\ref{Phi(ii)}),
\begin{align}
\Phi=&-C_S S^{(-)} l^2 \frac{\mathcal{H}}{a^2}
+\frac{1}{a}D^{(-)}
\Big( 
\frac{1}{a}
\big(\int d\eta a^2+C_{D}l^2\big)\Big)'\nonumber\\
&+O(k^{n_\Phi+2})\nonumber\\
=&-(C_S S^{(-)}+C_{D} D^{(-)})l^2\frac{\mathcal{H}}{a^2}
+D^{(-)}\frac{1}{a}\left(
\frac{1}{a}\int d\eta a^2\right)'\nonumber\\
&+O(k^{n_\Phi+2}).
\label{genPhi+}
\end{align}

The form of the fluctuation agrees with the form (\ref{u})
in the lowest order and thus confirms the absence
of the entropy perturbation after the bounce
in the long wavelength limit 
up to the order $k^{n_\Phi}$.
An immediate consequence of \eq{genPhi+} is that 
\be \label{DD}
D^{(+)} = D^{(-)}.
\ee 
This condition expressing the continuity of the curvature 
perturbation was shown
in \cite{Wands:2000dp} as the consequence of the
conservation law of energy momentum.
In this paper, we have 
established this as the result of the covariance of the matching 
condition. 

Note that our above result \eq{genPhi+} and \eq{DD} were 
obtained by assuming that in \eq{gconstphi} the term 
$k_i k_j S^{(-)} \cdots$  is higher order
in $k$ compared to the $D^{(-)}$ term. 
This is true for the  ekpyrotic scenario 
(see \eq{SD1} below), but not for the pre-big bang scenario \eq{SD0}. 
For a general potential of the form \eq{exppot}, 
one can check that the term $k_i k_j S^{(-)} \cdots$ 
will be dominant over the $D^{(-)}$ term 
whenever $1/3\leq p < 1$. Whenever the assumption is violated, 
one need to take into account of
the $k_i k_j S^{(-)}$ term in our above analysis 
and this leads to higher order corrections to
$M_{21}$ of the form $M_{21}\sim k^2$. 
And we obtain that
\be \label{DAS}
D^{(+)} = D^{(-)} + A k^2 S^{(-)},\quad \mbox{with $A$ a real number},
\ee 
in the leading order of $k$. This agrees with that of \cite{Brand}. 
In conclusion, the matching condition 
in the leading order
can be expressed in the form of a mixing matrix as follows
\begin{equation} 
\left(
\begin{array}{c}
S^{(+)}\\
D^{(+)}
\end{array}
\right)
= M \left(
\begin{array}{c}
S^{(-)}\\
D^{(-)}
\end{array}
\right)
=
\left(
\begin{array}{cc} 
C_S & C_D \\ 
A k^2 & 1
\end{array}
\right)
\left(
\begin{array}{c}
S^{(-)}\\
D^{(-)}
\end{array}
\right)
\label{general}
\end{equation}
with undetermined coefficients $C_S$ and $C_D$. 
\eq{general} is the most general 
matching condition 
that is consistent with  the requirement of local causality. 
This is one of the main result of this paper. 

Utilizing the explict spectra \eq{SD0} and \eq{SD1} below, the
condition \eq{DAS} implies that it is impossible to generate a scale
invariant spectrum for the pre-big bang and the ekpyrotic scenario. 
For the  general potential \eq{exppot}, a possibility of
obtaining a scale invariant spectrum was discussed in \cite{Brand} where
$p=2/3$ and $D^{(-)}\sim k^{3/2}$, $S^{(-)}\sim k^{-7/2}$  were found.
Unfortunately, the solution turned out not to be a
stable attractor point \cite{Heard:2002dr}.
It is still an open problem to obtain more general form
of initial spectrums which could result in the scale invariant
spectrum after the bounce.

Finally we remark  that if one instead uses the Lichnerowicz junction 
condition on the constant energy surface, one can apply the result of
\cite{DerMuk} and gets $C_S=1, C_D=0$.
Deviation of $C_S$ and $C_D$ from unity
and zero respectively is possible in general.
For example, in 
appendix \ref{sec:levelA_D}
we discuss the effect of having anisotropic stress during the bounce and 
how the values of $C_S$ and $C_D$ may be affected. 
We also remark that here we have only considered a single scalar 
coupled to gravity. For the
effect of multi-scalar scenarios on the matching condition, 
we refer the reader to appendix \ref{sec:levelA_C} for some discussions.

\subsection{\label{sec:level3_3}
Bouncing cosmologies and desnity spectrum}
 
In the studies of bouncing cosmologies, a potential is supposed to
be generated by the nonperturbative effects in string theory 
and it takes the form
\begin{equation}\label{exppot}
V(\varphi)=-V_0 e^{-\sqrt{\frac{6}{p}} l \varphi}.
\end{equation}
Different values of the parameter $p$ give different 
cosmological models.
For example, $0<p<<1$ in the ekpyrotic scenario, and
$p={\frac{1}{3}}$ in the pre-big bang scenario.  In this case, an exact
solution for the background metric (\ref{bmetric}) can be obtained. We have
\begin{equation}\label{scalefactor}
a=a_0 |\eta|^{\frac{p}{1-p}}
\end{equation}
and 
\begin{equation}
z=\left(\frac{3l^2}{2}\right)^{-1/2}a_0 p^{-1/2}|\eta|^{\frac{p}{1-p}},
\end{equation}
where $a_0$ is a parameter with the dimension of length. 
Moreover, the equation (\ref{schroedinger}) for the metric fluctuation
can be solved explicitly by 
\cite{Brand}
\begin{equation}
u=\left(\frac{3l^2}{2}\right)\sqrt{\frac{\pi}{2}}k^{-1}|\eta|^{1/2}
H_\nu^{(1)}(k|\eta|),
\end{equation}
where $H_\nu^{(1)}(k|\eta|)$ is the Hankel function of the first kind
and $\nu=\frac{1}{2}\frac{1+p}{1-p}$ .
Expanding the solution in $k$, we get
\begin{align}
S=&\left( \frac{3\pi l^2}{4} \right)^{1/2}\frac{2^\nu}{\sin{\nu\pi}}
\frac{a_0}{l^2}\frac{-i}{\Gamma(-\nu+1)}p^{-1/2}k^{-1-\nu}, \nn\\
D=&-\left( \frac{3\pi l^2}{4} \right)^{1/2}\frac{2^{-\nu}}{\sin{\nu\pi}}
a_0^{-1}\frac{-i e^{-i\nu\pi}}{\Gamma(\nu+1)}
p^{1/2}\frac{1+p}{1-p}k^{-1+\nu}.
\label{SandD}
\end{align}
These expressions give the $k$-dependence of the $S$ and $D$ for a given $p$. 
It is easy to obtain that
before the bounce, $S$ and $D$ scales as
\begin{align}
& S^{(-)}(k)\sim k^{-2}, \qquad D^{(-)}(k)\sim k^{0},\nonumber\\
&\mbox{for the pre-big bang scenario}, \label{SD0}\\
& S^{(-)}(k)\sim k^{-3/2}, \quad D^{(-)}(k)\sim k^{-1/2},
\nonumber\\
&\mbox{for the ekpyrotic scenario.}\label{SD1}
\end{align}
As has been argued 
above, scale invariant spectrum is not resulted
in these scenarios.

In conclusion, we have shown a no-go theorem: as long as the 
local causal matching condition is respected,
one cannot obtain a 
scale invariant spectrum in the post-bounce era
for both the pre-big bang scenario and the ekpyrotic scenario. This
no-go theorem can however be easily lifted.
In the next section we discuss the possibility of having 
nonlocal causality during the bounce and study the
possible effects on the resulting spectrum. In particular we demonstrate
that by allowing nonlocal effects during the bounce, a scale invariant
spectrum can be generated. 

\section{\label{sec:level4}
Non-local bounce and scale-invariant spectrum}

In string theory, noncommutative geometry is known
to be the possible and easily realized in terms of D-brane.
See for the review \cite{Douglas:2001ba}.
One possibility to violate the local causality condition 
is the emergence of noncommutative geometry.
Non-local causal structure is a general consequence of noncommutative
field theories \cite{Chu:2005nb}. 
In noncommutative space, it is known that the low-energy 
effective action typically includes nonlocal term.
For example in the noncommutative $\varphi^4$ theory 
the low energy effective action provides a nonlocal term 
\cite{Minwalla:1999px}
\begin{widetext}
\begin{equation}
\mathcal{L}_{\rm eff}
=\int d^4 p \frac{1}{2}
\left( p^2+m^2+\frac{g^2}{96\pi^2(p\circ p+\frac{1}{\Lambda^2})}\right)
\varphi(p)\varphi(-p)+\cdots ,
\label{Leff}
\end{equation}
\end{widetext}
where $p\circ p=|p_i (\theta^2)^{ij} p_j|$ and $\theta^{ij}$
is the noncommutative parameter with the relation
\begin{equation}
[ x^i,x^j ]=i\theta^{ij}.
\end{equation}
Here $x^i$ and $p_i$ are dimensionful quantities and 
correspond to $a x^i$ and $k_i/a$ in our paper.
It is clear from \eq{Leff} that the time evolution of $\varphi$ has an IR pole
if the ultraviolet cut-off scale $\Lambda$ is taken to  infinity.

We now consider the possible modification on the time evolution
of the metric fluctuation due to nonlocal effects.
Since it is not known how the gravity couple to the noncommutative
field theory, we will study a modified equation 
for the metric fluctuation as inspired by (\ref{Leff}). We assume that
within a time interval
$\eta_- <\eta < \eta_+$ around the bounce, nonlocal
effects become important and this is captured by the $C$-term 
in the following equation for the fluctuation
\begin{equation}
{u}''(k,\eta)+\Big(k^2+(C(\eta) k^{-\alpha})^2
- \frac{(z^{-1})''}{z^{-1}}\Big)u(k,\eta)
=0.
\label{ncshr0}
\end{equation}
Here $C(\eta)$ and $\alpha>0$ are dimensionless 
quantities. The time function $C(\eta)$ 
is taken to be nonzero in the time interval 
$\eta_- <\eta < \eta_+$. Moreover, as a simplification of our analysis, 
we assume that  within the bounce period $C(\eta)$ is 
constant and the $(C(\eta)k^{-\alpha})^2$ 
term dominates over the other two terms. 
This requires 
gravitational effects to be suppressed during the bounce.
We do not argue the mechanism of suppression and take the 
model as a toy model to demonstrate possible nonlocal
effects in the matching condition.

\begin{figure}[htbp]
\begin{center}
\includegraphics[scale=.7]{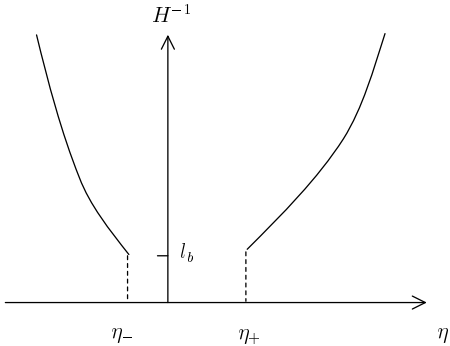}
\caption{\label{fig:2}The nonlocal bounce\label{Fig. 3}}
\end{center}
\end{figure}

We consider three regimes for the time evolution of the fluctuation. The
dividing line is the size of the Hubble scale. 
We assume that nonlocal effects starts to play a prominent role as 
the size of the Hubble scale gets down and becomes $H=1/l_b$. 
The length size $l_b$
characterizes the onset of new physics at sub-$l_b$ scale.
Therefore as the universe
approaches the bounce, its size shrinks rapidly until
$H =1/l_b$ at time 
$\eta=\eta_- <0$. 
Below that one can
no longer trust the semi-classical gravity picture, and the universe enters
into the nonlocal regime dictated by the term $(C(\eta) k^{-\alpha})^2$
in (\ref{ncshr0}) until $\eta=\eta_+$. At this moment, we again have
$H=1/l_b$ and the semi-classical gravity 
picture  is resumed. See FIG. \ref{fig:2}.

 Therefore,  outside the bounce period, we set 
$C(\eta)=0$, and \eq{ncshr0} reduces to
\begin{equation}
{u}''(k,\eta)- \frac{(z^{-1})''}{z^{-1}}u(k,\eta)
=0
\label{ncshr}
\end{equation}
for the superhorizon modes. Moreover, from now on, we will focus on the
ekpyrotic scenario with the type of scalar potential (\ref{exppot}),
i.e., $V(\varphi)=-V_0 e^{-\sqrt{\frac{6}{p}} l \varphi}$,
and we assume 
\begin{eqnarray}
p= \begin{cases}
0<q<<1 &\mbox{ if } \eta<\eta_-,\\
1/2 &\mbox{ if } \eta>\eta_+.
\end{cases}
\end{eqnarray}
The choices of $p$ correspond to the following picture in the ekpyrotic
scenario: before the bounce (i.e., $\eta<\eta_-$), the brane is slowly
approaching the bounce, and after the bounce (i.e., $\eta>\eta_+$) the
brane universe enters the radiation dominated era.
The time dependence of the scale factor outside the bounce period is
given by the semi-classical result given in section \ref{sec:level3}, 
and in summary 
\begin{eqnarray}\label{ab1}
a(\eta)= \begin{cases}
a_0 |\eta|^{\frac{q}{1-q}} &\mbox{ if } \eta<\eta_-,\\
b_0  \eta &\mbox{ if } \eta>\eta_+
\end{cases}
\end{eqnarray}
where $a_0, b_0$ are dimensionful length-scale constants. 
We should mention that for simplicity we have extrapolated the solution
\eq{ab1} so that $a(\eta=0)=0$. In general, one can relax this condition
by introducing the shift so that 
$a(\eta)=b_0(\eta-\eta_0)$ for
$\eta>\eta_+$. 
The analysis and the results are essentially the same. 
We will not consider this possibility.
As mentioned, the Hubble size serves as the dividing line at
$\eta=\eta_-$ and $\eta_+$, and these moments are characterized by
\be\label{Hubble}
H(\eta_-)=H(\eta_+)=l_b^{-1}.
\ee
Recall $\mathcal{H} := {\frac{p}{(1-p)\eta}}$, then 
from the above condition we obtain
\be\label{length}
\eta_+ = \sqrt{\frac{l_b}{b_0}} ,\quad \mbox{and} \quad 
\eta_- =  - q \frac{l_b}{a_0} .
\ee
It follows from (\ref{ab1}) and (\ref{Hubble}) that
\be\label{a-/a+}
\frac{a(\eta_-)}{a(\eta_+)}=q\frac{\eta_+}{|\eta_-|}.
\ee
We also introduce the ``period'' of the bounce  as 
\be
\Delta:= \eta_+ - \eta_-\approx\eta_+
\ee
since $q\ll 1$.

The fluctuation in each regime can then be solved easily and we have 
\begin{eqnarray}
u\approx \begin{cases}
S^{(-)} \frac{l^2 \sqrt{q}}{ A |\eta|^q}
+D^{(-)}
\frac{3 l^2 A  |\eta|^{1+q}}{ 2 \sqrt{q}} 
&\mbox{ for }\eta<\eta_-,\\
C_- e^{-i\tilde{k}\eta} +C_+e^{i\tilde{k}\eta}
&\mbox{ for }\eta_-<\eta<\eta_+,  \\
S^{(+)} \frac{l^2}{ \sqrt{2} B \eta} 
+D^{(+)} \frac{l^2B \eta^2}{ \sqrt{2}}
&\mbox{ for } \eta>\eta_+.\\
\end{cases}\label{solnnc}
\end{eqnarray}
Here we have denoted for simplicity
$\tilde{k}:=Ck^{-\alpha}$
and
\be
A := \left(\frac{3 l^2}{ 2}\right)^{-1/2} a_0,\qquad 
B := \left(\frac{3 l^2}{ 2}\right)^{-1/2} b_0.
\ee
Next we connect these solutions by requiring $u$ and $u'$ to be continuous
at $\eta=\eta_-$ and $\eta_+$. We have
\begin{align}
\left(
\begin{array}{c}
u\\
u'
\end{array}
\right)
&=
Z_{\eta=\eta_-}
\left(
\begin{array}{c}
S^{(-)}\\
D^{(-)}
\end{array}
\right)\nonumber\\
&:=
l^2 \left(
\begin{array}{cc}
\frac{\sqrt{q}}{ A  |\eta_-|^q} & \frac{3 A |\eta_-|}{ 2\sqrt{q}}\\ 
\frac{q^{\frac{3}{2}}}{ A |\eta_-|} & \frac{3 A |\eta_-|^q}{ 2\sqrt{q}}
\end{array}
\right)
\left(
\begin{array}{c}
S^{(-)}\\
D^{(-)}
\end{array}
\right)
\nonumber\\
&=
E_{\eta=\eta_-}
\left(
\begin{array}{c}
C_-\\
C_+
\end{array}
\right)\nonumber\\
&:=
\left(
\begin{array}{cc} 
e^{-i\tilde{k}\eta_-} & e^{+i\tilde{k}\eta_-} \\ 
-i\tilde{k} e^{-i\tilde{k}\eta_-} & i\tilde{k} e^{+i\tilde{k}\eta_-}
\end{array}
\right)
\left(
\begin{array}{c}
C_-\\
C_+
\end{array}
\right),
\end{align}
\begin{align}
\left(
\begin{array}{c}
u\\
u'
\end{array}
\right)
&=
Z_{\eta=\eta_+}
\left(
\begin{array}{c}
S^{(+)}\\
D^{(+)}
\end{array}
\right)\nonumber\\
&:=
l^2 \left(
\begin{array}{cc} 
\frac{1}{ \sqrt{2} B  \eta_+} &  \frac{B \eta_+^2}{ \sqrt{2}}\nonumber\\ 
-\frac{1}{ \sqrt{2} B \eta_+^2}
& \sqrt{2} B \eta_+
\end{array}
\right)
\left(
\begin{array}{c}
S^{(+)}\\
D^{(+)}
\end{array}
\right)\\
&=
E_{\eta=\eta_+}
\left(
\begin{array}{c}
C_-\\
C_+
\end{array}
\right)\nonumber\\
&:=
\left(
\begin{array}{cc} 
e^{-i\tilde{k}\eta_+} & e^{+i\tilde{k}\eta_+} \\ 
-i\tilde{k} e^{-i\tilde{k}\eta_+} & i\tilde{k} e^{+i\tilde{k}\eta_+}
\end{array}
\right)
\left(
\begin{array}{c}
C_-\\
C_+
\end{array}
\right).
\end{align}
Thus
\begin{align}
\left(
\begin{array}{c}
S^{(+)}\\
D^{(+)}
\end{array}
\right)
&= M \left(
\begin{array}{c}
S^{(-)}\\
D^{(-)}
\end{array}
\right)\nonumber\\
&=Z^{-1}_{\eta=\eta_+}E_{\eta=\eta_+}E_{\eta=\eta_-}^{-1}
Z_{\eta=\eta_-}
\left(
\begin{array}{c}
S^{(-)}\\
D^{(-)}\\
\end{array}
\right) 
\end{align}
We need nontrivial $M_{21}\sim O(k^0)$
to get the scale-invariant spectrum since $D^{(+)}\sim M_{21}S^{(-)}$
and $S^{(-)}\sim O(k^{-3/2})$.

 
The leading term in the small $q$
expansion of the matrix element $M_{21}$ is given by
\begin{align} \label{M21} 
M_{21}\approx &  \frac{\sqrt{2q}}{3\tilde{k}AB|\eta_-|\eta_+^2}
\Big(\tilde{k}(|\eta_-|+q\eta_+)\cos (\tilde{k}\eta_+)\nonumber\\
&+(-\tilde{k}^2\eta_+|\eta_-| +q)\sin (\tilde{k}\eta_+)\Big).
\end{align}
Now we have a nonzero $M_{21}$. 
To avoid the oscillating factor in (\ref{M21}),
we further assume that  
\be\label{keta+}
\tilde{k}\eta_+ \ll 1.
\ee
Then, 
\begin{align}\label{A21f}
M_{21}\approx &
\frac{\sqrt{2q}}{3AB|\eta_-|\eta_+^2}(|\eta_-|+2q\eta_+)\nonumber\\
=&\frac{\sqrt{2q}l^2}{2a_0b_0|\eta_-|\eta_+^2}
(|\eta_-|+2q\eta_+).
\end{align}
We obtain the scale invariant power spectrum
\begin{align} \label{Power}
P(k)&=k^3|D^{(+)}|^2=k^3|M_{21} S^{(-)}|^2\nonumber\\
&\sim
\left(\frac{l}{l_b}\right)^2
\left(2\frac{a(\eta_-)}{a(\eta_+)}+1\right)^2
\sim \left(\frac{l}{l_b}\right)^2.
\end{align}
Here we have used the explicit form of $S^{(-)}$ \eq{SandD}.
Since the observed CMB scale is about $10^{-5}$,
this implies that the Hubble scale around which the bounce 
occurs should be at around $10^{-5}$ of the Planckian energy scale.
This is quite reasonable. 
Since $S^{(-)}\sim k^{-\frac{3}{2}-q}$,
$q$ correction makes the spectrum slightly shifted to red.

Finally, we comment on the requirement coming 
from (\ref{keta+}). Now
\be \label{last1}
\tilde{k}\eta_+ =Ck^{-\alpha} \eta_+\ll 1.
\ee 
Since we are hoping to get the scale-invariant power spectrum in the
long wavelength limit, we only need this condition for very small
$k\eta_+$. The constraint (\ref{last1}) can be easily 
satisfied by requiring $0<\alpha <<1$ and $0<C \eta_+^{1+\alpha}<<1$. 
One may think that $\alpha<0$ also does the
job, but this will contradict to the fact that 
during the bounce period the term $Ck^{-\alpha}$ 
should dominate over the $k^2$ and the pumping
term 
in (\ref{ncshr0}) for the superhorizon modes. 
It is interesting to note that we will only need 
mild nonlocal effect (i.e., $\alpha<<1$)
instead of the canonical one in (\ref{Leff}) to have 
the scale-invariant spectrum. It deserves more efforts 
to derive such an effective theory
from string/M theory. 
 
In conclusion, we have shown that 
the nonlocal bounce and the corresponding 
matching condition is crucial in obtaining the scale-invariant 
CMB spectrum. For our calculation, we have assumed 
the equation for the fluctuation to be given by
(\ref{ncshr0}), which is based on a modified dispersion relation inspired by 
noncommutative field theory. 
It is  important to derive the precise form of 
the  nonlocality 
from a fundamental physical model such as string theory. 
We leave these issues for future study.
We hope our result helps to shed light on the problem of bouncing type
cosmologies by identifying the relevant kind of new physics 
that is needed.
 
\section{\label{sec:level5}
Conclusions and discussions}
In this paper, we have studied 
some possible effect of  physics 
beyond the general relativity on bouncing cosmologies.
Under the constraint of the local causality condition, 
we have derived the general form of the matching condition 
that relate the physics before and after
the bounce in the long wavelength limit. The possibility of mixing 
between the decaying mode and the constant mode is clarified. 
We find that the
coefficient of the decaying mode after the bounce can be
changed and can receive a correction coming from the constant mode. It
can happen even in the context of the general relativity if there exist
anisotropic stress during the phase transition. On the other hand, the
constant mode cannot be changed
if $k^2 S^{(-)}$ is sub-leading compared to $D^{(-)}$, as in the case of
the ekpyrotic scenario, for example. 
This conclusion is unaltered even considering effects coming from physics
beyond general relativity as long as the local causality condition 
is satisfied.
This rules out the possibility to obtain a scale-invariant spectrum
for the pre-big bang and for the ekpyrotic scenarios,
whenever local matching condition is employed. 
The identification of nonlocal effects as a possible outlet to 
achieve a scale invariant spectrum is one of the main results 
of this paper.
%

We have also studied 
the effects on the matching condition from  a violation of 
the local causality during the bounce.
With a toy model employing noncommutative geometry, 
we show that it is possible to obtain a scale invariant spectrum
if the nonlocal effects enter the evolution equation of the 
fluctuation as $k^{-\alpha}$ with $\alpha\ll 1$. 
Moreover,  from the CMB constraint, we find that the bounce should 
occur at 
a few  orders of magnitude below the Planck energy scale in our model.
The demonstration that a suitable form of nonlocal effects does lead to
a scale invariant spectrum is another main result of this paper.

It is an important task to understand the nature of spacetime 
in quantum gravity
and to derive more precisely the form of the nonlocal effects 
from string theory or other theory of quantum gravity. It will be very
interesting if the nonlocal effects we introduced in our toy model
do appear. We will leave
this for the future works.

Finally we comment on how nonlocal causality 
may play a role in the matching condition considered in other works.
\\
1. In \cite{KOSTpert}, it was argued that one can have a scale 
invariant spectrum by considering 
a linear matching condition
for $\epsilon_m=-\frac{2}{3} 
{\cal{H}}^2k^2\Phi$ and $\dot{\epsilon}_m$ at the bounce. 
It is easy to see that in order to  arrive at such a conclusion,
nonlocal causality  of the form we introduced in this paper is needed. 
This is because, as can be seen 
from \eq{def-scalar-fluct} and \eq{defPhi},
the definition of the gauge invariant quantity 
of $\Phi$ includes the spatial
integral of the 
fluctuations of the metric,
and hence  even if the matching condition
looks local in terms of $\Phi$ and $\dot{\Phi}$, 
it is not the case when the matching condition
is expressed in terms of the metric and the  scalar fields.
\\
2. In \cite{DurrVer}, negative surface tension was considered. By imposing 
a specific matching condition on the fluctutation of the surface tension, 
it was argued that one can obtain a scale invariant spectrum. Since the 
initial condition before the bounce is completely determined 
by the metric and the
scalar fields, thus in the framework of our general analysis, the 
proposed form of 
fluctuation of the surface tension with 
the choice of the matching surface 
should correspond to a violation of the 
local causality condition in order for a scale invariant spectrum
to be possible.
It will be interesting to have a concrete framework where the negative 
tension arises physically and to see how the violation of local causality 
comes about.
\\
3. In \cite{TTSpert} (see also \cite{Battefeld:2004mn}), 
matching conditions in the five dimensional 
context has been studied. 
In our analysis,  we are free to choose
the time of the matching $\eta_-$ so long as $k/aH(\eta_-)\ll 1$. 
The hypersurface
for the matching does not need to be the one on which
the bouncing phase start to happen. 
Since any fluctuation in the five dimensional framework can
be traced back to fluctuation in the four dimensional effective
theory if we go back in time. Thus, unless the time interval
during which the four-dimensional picture is not applicable is comparable
to the scale $k^{-1}$, 
we can choose to apply the matching condition 
in the four dimensional framework, and nonlocal causality is required
to obtain the scale invariant spectrum.
Finally, it would be interesting to see how the mechanism of mixing
of perturbation modes presented in the recent work \cite{McFadden:2005mq}
interconnect with the argument in this paper.
\begin{acknowledgments}

We would like to thank Takeo Inami for comments. 
We would like to thank the organizers of
the workshop "Summer School on String" held in Taipei, Taiwan
from 25 to 29 July, 2005 where the result of this paper was presented.
We also would like to thank the referee for important 
comments helping us to clarify some unclear points in the paper.
C.S. Chu would like to thank the Riken National Laboratory of Japan 
and the NCTS of Taiwan for hospitality.
K. Furuta would like to thank the Yukawa Institute for Theoretical 
Physics at Kyoto University. 
Discussions during the YITP workshop YITP-W-05-08 on 
"String Theory and Quantum Field Theory" were useful to complete 
this work. 
K. Furuta and F.L. Lin would like to thank VSOP for their hospitality 
in holding a workshop PLSS-06 "Physics for Large and Small Scales "  
at Hanoi for us to present and to finalize our work.
This work was partly supported by the EPSRC of UK and by the  
Taiwan's NSC grant 94-2112-M-003-014.

\end{acknowledgments}

\appendix
\section{\label{sec:levelA_A}
Metric and scalar fluctuation on the matching surface}

In this appendix, we show that one can exploit the residual gauge
symmetry of the syncronous gauge to choose the matching surface such
that $\varphi$ is constant over it. In this gauge $\delta g_{00}=\delta
g_{0i}=0$ and $\delta\varphi = 0$ at $\eta=\eta_-$. We will also 
compute the metric fluctuation \eq{gconstphi} around $\eta=\eta_-$ in
this gauge.

We start 
with the longitudinal gauge in which 
$B=E=0$. It also follows that $\phi=\psi=\Phi$ for perfect fluid.
Here $B,E,\phi,\psi$ are paremeters defined in 
\eq{def-scalar-fluct}.
In this gauge, the metric perturbation
and the scalar field fluctuation are given by
\begin{equation}
\delta g_{\mu\nu}=a^2
\begin{pmatrix} 
2\phi & 0 \\ 
0 & 2\phi \delta_{ij}
\end{pmatrix}
\end{equation}
and
\begin{equation}
\delta\varphi
=\left(\frac{3}{2}l^2\varphi'\right)^{-1}
\left(\phi'+\mathcal{H}\phi
\right).
\label{phivarphi1}
\end{equation}

Next, we perform the coordinate transformation
\begin{align}
&\eta^{(l)} \to \tilde{\eta}=\eta^{(l)} + \xi^{(l)0} (\eta^{(l)},\bx^{(l)})
\nonumber\\
&\mbox{ and }
x^{(l)i} \to \tilde{x}^i=x^{(l)i} +\delta^{ij}
\partial_j \xi^{(l)}(\eta^{(l)},\mbox{\boldmath{x}}^{(l)})
\label{gaugetr2}
\end{align}
to the gauge in which $\delta\varphi = 0$ and keeping $\delta g_{0i}=0$.
Here superscript $l$ represents
the longitudinal gauge. To do this, we need
\begin{equation}\label{xi0}
\xi^{(l)0}=-(\mathcal{H}'-\mathcal{H}^2)^{-1}(\phi'+\mathcal{H}\phi)
\end{equation}
and
\begin{equation}\label{xis}
\xi^{(l)}=\int d\eta \xi^0.
\end{equation}
To obtain \eq{xi0}, we have used 
the transformation law of the scalar quantity
\begin{equation}
\delta\varphi \to \delta\varphi+\varphi'\xi^0
\end{equation}
and
\begin{equation}
\varphi^{'2}=\frac{2}{3l^2}(\mathcal{H}^2-\mathcal{H}').
\end{equation}
Using \eq{flucttransf}, \eq{u-eq} and \eq{Phiin k}, 
the metric fluctuation in this gauge can be computed to give
\begin{widetext}
\begin{equation}
\delta g_{\mu\nu}^{(-)}(\tilde{\eta},\bk)
=2a^2\left(
\begin{array}{cc} 
\tilde{\phi} \;\;\;& 0 \\ 
0 \;\;\;& 
\begin{array}{c}\Big( D^{(-)}+O(k^{n_\Phi+2}))\Big) \delta_{ij}\\
+ k_i k_j \Big( S^{(-)}l^2\int d\eta a^{-2} 
- D^{(-)} \int d\eta(a^{-2} \int d\eta a^2)+O(k^{n_\Phi+2})
\Big)
\end{array}
\end{array}
\right)
\label{gconstphi2}
\end{equation}
\end{widetext}
Note that $\tilde{\phi}:=(2a^2)^{-1}\delta g^{(-)}_{00}=O(k^{n_\Phi+2})$.

To achieve   $\delta g^{(-)}_{00}=0$ to arrive synchronous gauge while
keeping $\delta\varphi=0$
on the surface of the matching,
we further perform the coordinate transformation 
\begin{equation}
\tilde{\eta} \to \eta=\tilde{\eta} + \tilde{\xi}^{0} 
(\tilde{\eta},\tilde{\bx})
\hspace{3mm}\mbox{ and }
\tilde{x}^i \to x^{i}=\tilde{x}^i+\delta^{ij}
\partial_j \tilde{\xi}(\tilde{\eta},\tilde{\mbox{\boldmath{x}}})
\label{gaugetr3}
\end{equation}
with 
\be\label{xi0new}
\tilde{\xi}^{0}=a^{-1}\int^{\tilde{\eta}}_{{\tilde{\eta}}_-}d \eta a
\tilde{\phi}
\ee
and
\be
\tilde{\xi}=\int^{\tilde{\eta}}_{{\tilde{\eta}}_-}d \eta 
\left( a^{-1}\int^{\tilde{\eta}}_{{\tilde{\eta}}_-}
d \eta a\tilde{\phi}\right).
\ee
Using \eq{flucttransf}, it is easy to obtain the metric 
and the scalar field fluctuation in this gauge,
\begin{widetext}
\be 
\delta g_{\mu\nu}^{(-)}(\eta,\bk)
=2a^2\left(
\begin{array}{cc} 
0 \;\;\;& 0 \\ 
0 \;\;\;& 
\begin{array}{c}\Big( D^{(-)}+O(k^{n_\Phi+2}))\Big) \delta_{ij}\\
+ k_i k_j \Big( S^{(-)}l^2\int d\eta a^{-2} 
- D^{(-)} \int d\eta(a^{-2} \int d\eta a^2)+O(k^{n_\Phi+2})
\Big)
\end{array}
\end{array}
\right)
\label{gconstphi3}
\ee
\end{widetext}
and 
\be
\delta\varphi(\eta,\bk)=\varphi' \tilde{\xi}^{0}
=O(k^{n_\Phi+2}).
\ee
Because of \eq{xi0new}, $\tilde{\xi}^0=0$ at 
$\tilde{\eta}=\tilde{\eta}_-$ and hence 
$\delta\varphi=0$ at $\eta=\tilde{\eta}_-$. 
In section \ref{sec:level3},   $\tilde{\eta}_-$ is denoted as $\eta_-$.

\section{\label{sec:levelA_B}
Comment on the choice of the matching surface}

In section \ref{sec:level3} we claim that one can 
choose any surface of the matching defined by
${\cal O}_-(\eta,\bx)={\cal O}^0_-=\mbox{constant}$ 
instead of the surface 
$\varphi(\eta,\bx)=\mbox{constant}$ without changing the 
matching condition at 
the order of $k$ we considered.
We will show this fact in this appendix.
For simplicity, we ommit 
the subscript $-$ and write ${\cal O}_-$ as ${\cal O}$.
First we will show that the leading power of $k$ dependence 
of the fluctuation $\delta {\cal O}(\eta,\bx)$ 
is $k^{n_\Phi+2}$
on $\varphi=\mbox{constant}$ surface.
To see this, consider possible scalar quantities
contributing to $\delta {\cal O}(\eta,\bx)$
by taking contraction of
spatial indices from the metric.
For example, these are
\begin{equation}
\delta(g'_{ij}g^{ij}) ,
\quad \delta g_{ij}k^i k^j,\cdots.
\label{possibilities}
\end{equation} 
Note that we do not consider contributions from
scalar field because $\varphi$ and $\varphi'$ 
has no fluctuation up to the order 
$k^{n_\Phi}$ 
in our gauge.
In (\ref{possibilities}) it is trivial that 
$\delta g_{ij}k^i k^j=O(k^{n_\Phi+2})$.
As for $\delta( g'_{ij} g^{ij})$, noting
$g_{ij}=a^2\delta_{ij}+2a^2\left(D \delta_{ij}
+O(k^{n_\Phi+2})\right)$ from (\ref{gconstphi}),
\begin{align}
g'_{ij} g^{ij}=&\left\{a^2\left( (1+2D)\delta_{ij}+O(k^{n_\Phi+2})\right)
\right\}' \nonumber\\
&\times a^{-2}\left( (1-2D)\delta_{ij}+O(k^{n_\Phi+2})\right)
\nonumber\\
=&6\mathcal{H}+O(k^{n_\Phi+2}).
\label{g'g}
\end{align}
Thus 
these terms contribute to $\delta {\cal O} (\eta,\bx)$ 
at order $O(k^{n_\Phi+2})$.
Similarly one can
verify this for any other quantity and
thus $\delta {\cal O}(\eta,\bx)$ 
is suppressed with a 
factor of $k^2$ compared to $\Phi$. 
This is similar to the
suppression of the entropy perturbation.

Next, we change the surface of the matching 
to one defined by 
${\cal O}(\eta,\bx)=\mbox{constant}$ and show that it only
has an effect on the higher order terms in the matching condition.
To change the surface of the matching, we 
perform the coordinate transformation
\begin{equation}
\eta \to \tilde{\eta}=\eta+\xi^0(\eta,\bx)
\end{equation}
and consider the matching condition on 
$\tilde{\eta}=\tilde{\eta}_-$ surface.
Demanding $\delta{\cal O}\to \delta{\tilde{\cal O}}=
\delta{\cal O}+{\cal O}'\xi^0(\eta_-,\bx)=0$,
we get
\begin{equation} \label{xi0a}
\xi^0|_{\eta=\eta_-}=-\frac{\delta{\cal O}}{{\cal O}'}\Big|_{\eta=\eta_-}
=O(k^{n_\Phi +2})
\end{equation}
since $\delta{\cal O}=O(k^{n_\Phi +2})$.
To satisfy the syncronous gauge condition 
$\delta g_{00}=\delta g_{0i}=0$,
one needs
\begin{equation}
\tilde{\phi}=\phi-(a'/a)\xi^0-(\xi^0)'=0
\label{ap1phi}
\end{equation}
and 
\begin{equation}
\tilde{B}=B+\xi^0-\xi'=0.
\label{ap1B}
\end{equation}
$\phi$ and $B$ are zero as is read off from (\ref{gconstphi}) and 
$\xi$ is the parameter for the diffeomorphism in the
spatial direction
\begin{equation}
x^i\to x^i +\delta^{ij}\partial_j \xi(\eta,x).
\end{equation}
(\ref{ap1phi}) and (\ref{ap1B}) are solved by
\begin{equation}
\xi^0=C a^{-1} 
\end{equation}
and 
\begin{equation}
\xi=C\int d\eta a^{-1},
\end{equation}
where $C=-a\frac{\delta {\cal O}}{{\cal O}'}|_{\eta=\eta_-}=O(k^{n_\Phi+2})$ 
due to \eq{xi0a}.
Substituting these parameters in (\ref{flucttransf}),
one obtains
\begin{equation}
\delta g_{ij}\to\delta \tilde{g}_{ij}=\delta g_{ij} +2a^2\left(
C\frac{a'}{a^2}\delta_{ij}+\int d\eta a^{-1} C|_{ij}
\right),
\end{equation}
where $\delta g_{ij}$ in the RHS is given by (\ref{gconstphi}).
Thus  
a different choice of the matching surface 
only affects
the higher order terms in the fluctuation of the metric. It is also easy to
verify that $\delta \varphi=O(k^{n_\Phi+2})$ and $\delta \varphi'
=O(k^{n_\Phi+2})$
on the $\tilde{\eta}=\tilde{\eta}_-$ surface. 
Therefore the fluctuations on 
the ${\cal O}=\mbox{constant}$
surface take the same form as (\ref{gconstphi})
and (\ref{phiconstphi}) up to the order we consider
and hence the same matching condition (\ref{general}) is resulted.


\section{\label{sec:levelA_C}
Effects of entropy perturbation}

In this appendix we comment on the entropy perturbation and its effect
in modifying the matching conditions. 

Entropy perturbation is the perturbation seen by a local observer and
can affect the local matching condition.
Entropy perturbation is defined by (see e.g., \cite{Wands:2000dp})
\begin{equation}
\mathcal{S}=a\mathcal{H}\left(\frac{\delta p}{p'}-
\frac{\delta\rho}{\rho'}\right)
\label{entropypert}
\end{equation}
where $\rho$ and $p$ are energy density and pressure
respectively. 
If there is no entropy perturbation, (\ref{entropypert}) 
indicates that the fluctuation 
of pressure and the energy density 
can be reinterpreted as a
local time delay. Thus for the purely 
adiabatic perturbation,
every portion of the space experiences the same history.

For a single scalar field model as considered in this paper, entropy
perturbation is known to be $k^2$ order higher than $\Phi$
\cite{Bassett:1999mt}. This is the reason why the entropy fluctuation
can be ignored in our analysis. 

For multi-fields case, one can
decompose the fields into adiabatic field and entropy fields
\cite{Gordon:2000hv}. In the Minkowski space regime, entropy
perturbations are $k^1$ order higher than $\Phi$. 
If the bounce occurs
in a time period much shorter than the wavelength as in the ekpyrotic
scenario with $p\ll 1$, one only has 
modification at the order of
$k^{n_\Phi+1}$ in the matching condition (\ref{general}). 
There is no
modification in the lowest order and thus 
one cannot obtain a scale invariant mode from $D^{(+)}$.
On the other hand, if there is enough time comparable to $k^{-1}$ prior
to the bounce, one can change the order of
$D^{(-)}$. 
Indeed, changing the
parameter $p$ in the scalar potential will change the power of $k$ of
$D^{(-)}$. 
There have been
attempts to get scale invariant spectrum by having 
sufficient 
duration in the
contracting phase \cite{Finelli:2001sr,DiMarco:2002eb}. It was argued
in 
\cite{Brand,Finelli:2001sr} 
that a 
scenario with $p=2/3$ is able to provide a
scale invariant spectrum. Unfortunately the solution turned out 
not to be a stable attractor point \cite{Heard:2002dr}.

\section{\label{sec:levelA_D}
Effects of anisotropic stress}

To see the effect of anisotropic stress, we assume
the validity of the general relativity during the bounce and 
consider the 
metric fluctuation of the form (in the syncronous gauge)
\begin{equation}
\delta g_{\mu\nu}=2a^2 
\begin{pmatrix} 
0 & 0 \\ 
0 & D\delta_{ij}
+F_{ij}
\end{pmatrix}
\end{equation}
with the initial condition given by (\ref{gconstphi})
at $\eta=\eta_-$. The Einstein tensor for $i\not=j$ 
up to the order 
$k^{n_\Phi}$ becomes
\begin{equation}
G^i_j=a^{-2}\left(k_i k_j D(x)
+F_{ij}''+2\mathcal{H}F_{ij}'\right).
\end{equation}
Note the indices are not raised in RHS.
Einstein equation $G^i_j=3 l^2 T^i_j$ can be solved 
by
\begin{equation}
F_{ij}=A_{ij}\;l^2 \int d\eta  a^{-2}-k_i k_j D
\int d\eta \left(a^{-2}\int d\eta a^2\right)
+B_{ij}(\eta)
\label{solF}
\end{equation}
with constants $A_{ij}$ and
\begin{equation}
B_{ij}(\eta)=3l^2\int d\eta\left( a^{-2}
\int d\eta \; a^4 T^i_j \right).
\label{deltaS}
\end{equation}

Initial condition (\ref{gconstphi}) can be satisfied
by setting $A_{ij}=k_i k_j S^{(-)}$.
Since $T^i_j\not=0$ only in the region $\eta_-<\eta<\eta_+$,  
after $\eta=\eta_+$ (\ref{solF}) becomes 
\begin{align}
F_{ij}=&\Big( k_i k_j S^{(-)}+3\int_{\eta_-}^{\eta_+}
d\eta \; a^4 T^i_j \Big)  
l^2 \int d\eta  a^{-2}\nonumber\\
&-k_i k_j D
\int d\eta \left(a^{-2}\int d\eta a^2\right) .
\label{solF2}
\end{align}
This results in a jump in $S$:
\begin{equation}
k_i k_j S^{(+)}=k_i k_j S^{(-)}+3\int_{\eta_-}^{\eta_+}
d\eta \; a^4 T^i_j.
\end{equation}
Since the background stress tensor is
diagonal, the off-diagonal components of $T^i_j$ come from the
fluctuation and should be linear in $S^{(-)}$ or 
$D^{(-)}$.  Thus the contribution to $S^{(+)}$ is also
linear in $S^{(-)}$ or in $D^{(-)}$ as in (\ref{general}) but the
details of the mixing depends on the physics during the bounce. 

In conclusion, after taking into account the anisotropic stress
during the bounce, the matching condition (\ref{general}) 
remains the most
general one in the framework of general relativity. 
Physics beyond general relativity can 
only affect $C_S$
and $C_D$ quantitatively, and there is no contribution from $S^{(-)}$ to
$D^{(+)}$ 
if local causality is assumed. 


\end{document}